\def\aa{{A\&A}}

\def\aj{{AJ}}
\def\annrev{{ARA\&A}}
\def\apj{{ApJ}}

\def\mnras{{MNRAS}}
\def\nat{{Nature}}

\def\farcs{\hbox{$.\!\!^{\prime\prime}$}}
\def\mbh{M_\bullet}

\documentclass[cup5b]{caps}
\usepackage{graphicx}
\usepackage{amssymb}
\usepackage{ociwsymp1}
\HeadText{P. T. de Zeeuw}

\begin{document}

\pagenumbering{arabic}

\author[]{P. TIM DE ZEEUW\\Leiden Observatory}

\chapter{Conference Summary}

\section{Introduction}

In the past decade, much effort was devoted to measure the masses of
supermassive black holes in galactic nuclei, to establish the relation
between black hole mass and the global/nuclear properties of the host
galaxy, and to understand the role of these objects in the formation
and subsequent dynamical evolution of galaxies. This whole area of
research is an appropriate and timely topic for the first in this
series of symposia. I congratulate the organizers, and in particular
Luis Ho, for putting together such an interesting program of talks
and posters.

Nearly half the contributions were devoted to the various methods that
are being used to estimate masses of central black holes, from nearby
globular clusters all the way to quasars at redshifts $z> 6$.  This
work forms the foundation of our understanding of black hole
demography, and provides constraints on scenarios for black hole
formation and growth, and on their connection to the formation and
properties of the host galaxy. I will address these topics in this
same order.

\section{Black Hole Masses}

The basic principle is to deduce the mass distribution in a galactic
nucleus from the observed motions of stars or gas, and compare it with
the luminous mass associated with the observed surface brightness
distribution. If this requires a central mass-to-light ratio $M/L$
larger than anything that can be produced by normal (dynamical)
processes, the associated dark mass is considered to be a black hole.

Depending on the nature of the nucleus, and on its distance, one may
employ the kinematics of individual stars, the spatially-resolved
kinematics of the absorption lines of the integrated stellar light, or
of the emission lines of gas (optical/maser), or the unresolved
kinematics from reverberation mapping and modeling of line profiles.
It is useful to review the contributions on all these approaches by
starting with the nearest black holes, and working to larger
distances. This takes us from what could be called ``primary'' black
hole mass indicators to ``secondary'' and ``tertiary''
estimators.\looseness=-2

\subsection{The Galactic Center}

The nearest supermassive black hole is located at the Galactic
Center. Despite the large foreground extinction, it is possible to
detect individual stars near it in the infrared, and to measure their
proper motions and radial velocities. Ghez showed us beautiful results
based on adaptive-optics-assisted near-IR imaging with a spatial
resolution of $\sim$$0\farcs05$. The velocities of stars this close to
the Center are large: the five-year coverage to date already defines
accurate three-dimensional stellar orbits for at least six stars, some
of which are very eccentric. The stars appear to have early-type
spectra, so may be massive and young. The current record holder
recently passed pericenter at only 60 AU from the black hole, with a
velocity of $\sim$9000~km~s$^{-1}$!

With data of this quality, measuring the mass at the Galactic Center
reduces to the classical astronomy of binary orbits. The current best
estimate is $3.6\times 10^6 M_\odot$. The spatial resolution achieved
allows probing well inside the radius of influence of the black
hole. The Galactic Center hence provides a unique laboratory for
measuring the properties of a star cluster in the regime where the
black hole dominates the gravitational potential, and allows studying
the formation and evolution of massive stars under atypical conditions
(see contributions by Alexander and Scoville).

The generally friendly but nevertheless intense rivalry between the
groups at Keck (Ghez) and at ESO (Genzel/Eckart) promises further
interesting results in the near future. In particular, the near-IR
integral-field spectrographs {\tt SINFONI} and {\tt OSIRIS}, and
hopefully {\tt NIFS} on Gemini, will provide proper motions and radial
velocities from the same observations, and will push these studies to
fainter magnitudes.

\subsection{Globular clusters}

Measurement of the resolved stellar kinematics is also possible in
nearby globular clusters, which might contain ``intermediate-mass''
black holes, i.e., with masses in the range $10^3-10^4 M_\odot$.
Ground-based proper motion and radial velocity studies generally lack
the spatial resolution to observe the kinematic signature of such
black holes, but the {\it Hubble Space Telescope (HST)}\ can detect them.

Van der Marel discussed the recent measurement of radial velocities in
the central arcsecond of M15 obtained with {\tt STIS}.  The
interpretation of the observed radial velocity distribution has led to
significant controversy, which centers around the nature of the
measured inward increase of $M/L$. Because dynamical relaxation has
been significant in M15, dark remnants must have aggregated in its
center. The initial estimate of a black hole of $\sim$$3\times 10^3\,
M_\odot$ had to be revised because of errors in previously published
figures describing a Fokker-Planck model for the cluster.  The
revised figures suggest that the observed $M/L$ increase could be due
entirely to dark remnants, but leaves room for a black hole of
$\sim$$10^3 \,M_\odot$.  Whatever the final word, this work suggests
that it might be profitable to scrutinize other globular clusters, and
it pushes the groups that have been producing detailed dynamical
models for globular clusters to provide them for specific objects,
while fitting all kinematic data.\looseness=-2

{\it HST}\ proper motions might also reveal central black holes in globular
clusters. The case of $\omega$ Centauri is of particular interest. It
is a massive cluster that may be the remnant of a galaxy that fell in
long ago, shows little evidence of mass segregation, and has
ground-based kinematics for $\sim$$10^4$ stars. It is the target of an
extensive {\tt WFPC2} proper motion study (King \& Anderson 2002)
which should detect a black hole if it has a mass as large as expected
from the $\mbh-\sigma$ relation for galaxies (see below).

\vfill\eject
\subsection{Stellar absorption-line kinematics}

We cannot resolve motions of individual stars in the nuclei of other
galaxies, but it is possible to measure the stellar kinematics from
the integrated light at high spatial resolution. Ground-based
integral-field spectroscopy is crucial for constraining the stellar
$M/L$ and the intrinsic shape of the galaxy. The orbital structure in
these systems is rich, so a true inward increase of $M/L$ must be
distinguished from a possible radial variation of the velocity
anisotropy. This requires measuring the shape of the line-of-sight
velocity distribution as a function of position on the sky, and then
fitting these with dynamical models. Recent work in this area was
reviewed by Gebhardt and Richstone, with assists by Kormendy and
Lauer.

A number of independent codes have been developed to construct
dynamical models with axisymmetric geometry. These are by now all
based on Schwarzschild's (1979) numerical orbit superposition method
(Rix et al.\ 1997; van der Marel et al.\ 1998). The consensus view is
that they give consistent black hole masses, as long as one is careful
to obtain smooth solutions (by maximum entropy or via regularization,
or by including very large numbers of orbits). A case study is
provided by M32, which has been modeled by many groups over the past
20 years. The latest analysis is by Verolme et al.\ (2002), who
model major-axis {\tt STIS} data as well as {\tt SAURON}
integral-field kinematics. This produces an accurate $M/L$, for the
first time fixes the inclination of M32, and sets the mass of its
black hole at $2.5\times 10^6 M_\odot$.

It is useful to keep in mind that two other nearby nuclei differ
considerably from that in M32. The nucleus of M31 has long been known
to be asymmetric (Light, Danielson \& Schwarzschild 1974). {\tt TIGER}
and {\tt OASIS} integral-field spectroscopy revealed its stellar
kinematics (Bacon et al.\ 2001), and allowed understanding of previous
long-slit work, including {\tt FOS}, {\tt FOC} and {\tt STIS} data
(Kormendy, Emsellem). Long-slit absorption-line spectroscopy of the
nucleus of the Sc galaxy M33 shows that it cannot contain a black hole
more massive than $1.5\times 10^3 M_\odot$ (Gebhardt et al. 2001).

To date, nearly 20 early-type galaxies with distances up to 20 Mpc
have been studied in this way. Dynamical modeling of ground-based and
{\tt STIS} spectroscopy provides evidence for black holes in all of
them. More than half the determinations come from the long-running {\it
HST}\
program by the Nuker team, and are based on edge-on axisymmetric
models (Gebhardt et al.\ 2003). Additional cases can be found in the
poster contributions (e.g., Cappellari, Cretton). The black hole
masses usually lie in the range $5\times 10^{7-8} M_\odot$, and may be
accurate to 30\% on average.  The inferred internal orbital structure
suggests tangential anisotropy near the black hole. This constrains
black hole formation scenarios (Sigurdsson, Burkert).  However, many
of these objects display kinematic signatures of triaxiality, and
surely not all of them are edge-on. This casts doubt on the accuracy
of the individual masses. Application of the same modeling approach,
but now for triaxial geometry, will show how severe these biases
are. The machinery for doing this is now available (Verolme et al.\
2003).\looseness=-2

Detection of stellar kinematic evidence for black holes smaller than
$5\times 10^7 M_\odot$ in galaxies at the typical distance of Virgo
requires higher spatial resolution than {\it HST}\ offers. Here other
methods
will be needed. The central surface brightness in the giant
ellipticals is too low for the 2.4-m {\it HST}\ to provide spectroscopy of
sufficient signal-to-noise ratio at high spatial resolution. They are
prime targets for near-infrared integral field spectroscopy of the CO
bandhead at 2.3 $\mu$m, with 8-m class telescopes.

\subsection{Optical emission lines}

The nuclei of active early-type galaxies, and those of most spirals,
contain optical emission-line gas. Its kinematics can be used to
constrain the central mass distribution, as long as the spatial
resolution is sufficient to resolve the region where the gravity field
of the black hole dominates the motions. It is important to account
for the proper observational set-up. Maciejewski showed that not doing
so can change an inferred black hole mass by a factor as large as
4! It is also important to confirm that the gas is in regular
motion, by using an integral-field spectrograph, or by taking multiple
parallel slits.

To date, the emission-line gas in about 100 nuclei has been observed
with {\it HST}, for galaxies to distances of up to 100 Mpc (see
contributions by Axon, Barth, Marconi, Sarzi, Verdoes Kleijn).  Only
$\sim$20\% of these seem to have circular rotation, and so far only a
few black hole masses derived in this way have been published, all of
them larger than $\sim$$10^7 \,M_\odot$. NGC~3245 is a textbook example
(Barth et al.\ 2001).

The accuracy of these black hole masses is not easy to establish.  To
date, there are very few cases for which a black hole mass has been
determined by more than one independent method. While good results are
claimed for NGC 4258 (stellar kinematics/maser emission --- see below),
the mass derived for the black hole in IC 1459 from gas motions is
significantly lower than that derived from stellar motions. It will be
very useful to carry out such comparisons for other nuclei, as this is
the only way to put the masses determined by different methods onto
the same scale, a prerequisite for an unbiased analysis of the black
hole demography. It seems clear that the kinematic model of circular
rotation for the gas is nearly always too simple. In many cases the
gas velocity dispersion increases strongly inward. This is evidence
for ``turbulence,'' presumably caused by nongravitational motions, which
probably include in/outflows, winds, and genuine turbulence in the gas
near the central monster (similar to what is seen in the simulations
reported by Wada). An improved (hydrodynamic) standard model for
analysis of the gas motions would be very useful.\looseness=-2

\subsection{Masers in Seyferts and LINERs}

VLBI measurements of H$_2$O maser emission allow probing the gas
kinematics to very small scales, and have provided the cleanest
extragalactic black hole mass determination, for NGC~4258. This
method achieves the highest spatial resolution to date, but is
possible only when the circumnuclear disk is nearly edge-on. A search
of $\sim$700 nuclei has revealed maser emission in 21 additional
cases, and regular kinematics on VLBI scales in four of these, to
distances of up to 70~Mpc. The inferred black hole masses range from
$(1-40)\times 10^6 M_\odot$.  This mass range is currently very
difficult to probe otherwise.

\subsection{Reverberation mapping}

The observed time-variation of broad emission lines (e.g., H$\beta$)
relative to the continuum in nearby Seyfert 1 nuclei and quasars can be used
to
estimate the radius of the spatially unresolved broad-line region by means
of
the light travel-time argument. In combination with simple kinematic
models for the motion of the broad-line clouds, the observed width of
the lines then provides a typical dispersion, so that an estimate of
the mass of the central object responsible for these motions follows
(Green, Peterson).

This is a powerful method in principle, as it allows probing nuclei out
to significant distances. Early mass determinations appeared to be
statistically systematically lower than those obtained by other means,
but this discrepancy has now disappeared. However, there is not yet a
truly independent calibration of the mass scale. The nearest
well-studied case is too faint for obtaining, e.g., the resolved
stellar kinematics of the nucleus, and for this reason the simple
kinematic model is calibrated by requiring results to fit the local
$\mbh-\sigma$  relation, so that the masses agree by
definition.  This makes reverberation mapping effectively a
``secondary'' mass indicator.

\subsection{Line widths}

A number of contributions (e.g., Vestergaard, Kukula, Jarvis
\& McClure) advocated using the widths of the H$\beta$, Mg~II,
or C~IV emission lines to estimate a velocity dispersion $\sigma$, and
to infer a typical radius $R$ from a locally observed correlation between
luminosity and broad-line region size, so
that a black hole mass follows. These authors set the overall scale by
calibrating versus reverberation mapping masses, which in turn is
calibrated by the local $\mbh-\sigma$  relation, making this a
tertiary mass indicator, which may still contain significant
systematic uncertainties.  This approach is potentially very useful,
as it can be applied even to the highest redshift quasars. The method
suggests that some of the $z>6$ quasars may contain $5\times 10^9
M_\odot$ black holes. If true, this would require very rapid growth of
any seed black hole to acquire so much mass at such early times. It
will be very useful to calibrate this method more directly.

\subsection{X-ray emission}

Recently, it has become possible to measure the profile of the Fe
K$\alpha$ line at 6.4 keV in the nuclei of Seyfert galaxies such as
MCG-6-30-15 and NGC 3516. The early {\it ASCA} results have now been
superseded by {\it XMM-Newton}\ spectra (Fabian et al.\ 2002).  The data
reduction effort is significant, but the asymmetric line profile shows
a width of about $10^5$~km~s$^{-1}$, a sure sign that the emission must
originate near 10 Schwarzschild radii of a (rotating) relativistic
object. The radius where the emission originates is not measured
independently, so these observations do not provide the mass of the
black hole.

\section{Black Hole Demography}

\subsection{$\mbh-\sigma$ relation}

Early attempts to relate measured black hole masses to global
properties of the host galaxy focused on $\mbh$ versus bulge
luminosity $L_B$ (e.g., Kormendy \& Richstone 1995). Following a
suggestion by Avi Loeb, two groups showed that there is a tighter
correlation between $\mbh$ and host galaxy velocity dispersion
$\sigma$ (Ferrarese \& Merritt 2000; Gebhardt et al.\ 2000). This
generated many follow-up papers either explaining the relation, or
``predicting'' it afterwards, and also sparked a remarkable (and
sometimes even amusing) debate about the slope of the correlation, and
the proper way to measure it (see Tremaine et al.\ 2002 for a recent
summary, and contributions by Gebhardt, Kormendy, and
Richstone).\looseness=-2

We have seen that the black hole masses that form the foundation of
the $\mbh-\sigma$ relation may still contain significant
errors. The black hole masses for high-$\sigma$ galaxies (giant
ellipticals) generally rely on gas kinematics, and these should be
treated with caution until we have independent measurements derived
from stellar kinematics. Many of the available stellar dynamical
measurements in E/S0 galaxies cover black hole masses in the range
$5\times 10^7$ to $5\times 10^8 M_\odot$, but rely mostly on edge-on
axisymmetric dynamical models. Many of the host galaxies show signs of
triaxiality, which must cause considerable changes in the anisotropy
of the orbital structure relative to that in an axisymmetric
model. Depending on the direction of viewing, one might observe more
or less of this anisotropy in the line-of-sight kinematics, and
accordingly ascribe less or more of a higher central velocity
dispersion to the presence of a black hole. This may be a significant
effect for any given galaxy (Gerhard 1988). While the resulting shifts
in black hole mass may tend to average out when observing a sample of
galaxies with random orientations, there may still be residual
systematic errors in $\mbh$ when derived with an axisymmetric
model. If the distribution of intrinsic shapes of early-type galaxies
varies with luminosity (or $\sigma$) then relying on axisymmetric
models may introduce a bias in the slope of the $\mbh-\sigma$
relation.

The number of reliable black hole masses below $4 \times 10^7 M_\odot$
is still modest, especially for spiral galaxies. The expectation is
that the spirals with classical spheroidal bulges should all contain a
black hole, in agreement with the fact that a very large fraction of
them also contains a low-luminosity active nucleus (Heckman, Ho,
Sadler). The smallest reported black hole masses for spheroidal
systems are those for the Galactic globular cluster M15 and for the
M31 companion cluster G1. While both are consistent with an
extrapolation of the $\mbh-\sigma$ relation to small $\sigma$, the
mass determinations are not yet secure, in particular for
M15.\looseness=-2

In view of this, it seems to me premature to consider the $\mbh-\sigma$
relation ironclad over the entire range of $10^3-3\times
10^9 M_\odot$, {\it and} valid for all Hubble types to all redshifts,
as some speakers assumed, and as appears to have become almost
universally accepted outside the immediate field.  Rather than
debating the details of finding the slope of the correlation, it seems
more productive to measure more reliable black hole masses, and indeed
to work out why this correlation should be there, especially if it
were to extend all the way from giant ellipticals to globular
clusters.

\subsection{Galaxies with pseudo-bulges}

Many spiral galaxies do not have a classical $R^{1/4}$ spheroid, but
instead contain a separate central component that has many properties
resembling those of disks (Kormendy 1993; Carollo 1999), and is
commonly referred to as a pseudo-bulge.  Carollo reviewed recent work
on these objects, based on large imaging surveys with {\it HST}. She
showed that all pseudo-bulges in her sample contain nuclear star
clusters that are barely resolved at {\it HST}\ resolution (e.g.,
Carollo et al.\ 2002). The example of M33 mentioned in the above
suggests that these galaxies may not have a central black hole,
consistent with the almost complete absence of emission-line
signatures of activity (Heckman, Ho). The sole exception appears to be
NGC~4945 which has a black hole mass based on maser emission.  It will
be very interesting to establish what the demography of black holes is
in galaxies with pseudo-bulges.  The presence of the tight central
cluster will make it very difficult to measure central black hole
masses by stellar dynamical methods.

\subsection{Quasars}

We heard much about recent work on quasars, driven by new and large
samples, notably that provided by the Sloan Digital Sky Survey.  This
has produced accurate and detailed luminosity functions (Osmer, Fan),
and also tightened the limits on the typical lifetime of the quasar
phenomenon derived from the spectacular decline in numbers at
redshifts below two, to $\sim4\times 10^7$ yr (Martini).

The $z> 6$ quasars seen in the Sloan Survey are interesting objects
for a number of reasons. The most luminous of these demonstrate that
the central black hole responsible for the observed activity must have
grown about as fast as possible. The spectrum shortward of Ly$\alpha$
also provides evidence for a Gunn-Peterson trough, suggesting the
tail end of reionization occurred a little above a redshift of
six.

Many apparently conflicting opinions were stated on the nature of the
host galaxies of quasars. Lacy, Kukula and Haehnelt argued that the
high-$z$ quasars are located in (boxy) ellipticals. Peng reported a
mix of host galaxy types at $z\approx 1$, Heckman argued for ``very
luminous galaxies,'' and Scoville presented evidence that many of the
nearest low-luminosity quasars reside in disks. This underlines the
importance of careful sample definition and nomenclature before
drawing strong conclusions on host-galaxy evolution with redshift.

A number of speakers revisited  So\l tan's (1982) line of reasoning, and
used recent data to work out the mean density in black holes versus
the mean density in extragalactic background light. Fall derived that
the quasar contribution is 1\% of that of stars. The consensus value
for the mean density of black holes is $\sim2.5\times 10^5\,
M_\odot$ Mpc$^{-3}$.  This assumes an efficiency of about 0.1, and agrees
with the local census in average value. However, the comparison of the
quasar luminosity function and the current estimate of the local black
hole mass function is less straightforward, and requires more work
(Yu, Richstone).\looseness=-2

Fabian addressed the possibility that the recent measurements of the
X-ray background with {\it Chandra}\ and {\it XMM-Newton}\ indicate a
black hole density a factor of 4 larger than the above estimate. This
would either require a much-increased accretion efficiency (which
seems hardly possible), or over two-thirds of the active nuclei being
obscured. He however also presented evidence that the X-ray objects
evolve differently with redshift than do the quasars, and in the end
concluded (or, as some pointed out, recanted) that any discrepancy
might be less than a factor of 2. In my primitive understanding of
this issue this means that we are close to having the correct census
of black holes in the Universe.

\section{Black Hole Formation and Growth}

Rees (1984) published a celebrated diagram illustrating that the
formation of (seed) black holes early on during galaxy formation is
essentially inevitable. Many of the processes he discussed were
considered also during this conference.

\subsection{Formation}

Recent work on the formation of the first objects in the Universe
suggests that the initial collapses occurred as early as $z\approx 20$.
Whether these lead to mini-galaxies or mini-quasars is less clear
(Haiman), but either way, the reionization of the Universe must have
started about this time. The theorists seemed in agreement that most
of the mass in the black holes that power the active nuclei must have
grown from seeds formed around $z\approx 15$ (Phinney). Clarke argued for
competitive accretion (by analogy with the normal star formation
process), while others considered the collapse of a supermassive star
to a Kerr black hole (Shapiro) and the dynamical evolution of dense
stellar clusters (Freitag, Rasio). It is clear from the existence of
the $z\approx 6$ quasars that whatever happened, the process was able to
build black holes with masses larger than $10^9 M_\odot$ very quickly.

\subsection{Adiabatic growth}

A number of speakers considered processes that cause (secular) growth
and evolution of a central black hole, with as ultimate aim to deduce
(or at least constrain) the formation history of the black hole from
the observed morphology and kinematics of the host nucleus. Sigurdsson
considered the problem of adiabatic growth through accretion of
stars. This process is well understood in spherical geometry, and
produces a power-law density profile with a range of slopes, as well
as tangential orbital anisotropy near the black hole. However, the
same observed properties can be produced by other formation
mechanisms, such as violent relaxation around a pre-existing black
hole (Stiavelli 1998). The problem of adiabatic growth in axisymmetric
or triaxial geometry deserves more attention. There is a fairly
wide-spread expectation of inside-out evolution of the shape toward a
more nearly spherical geometry, but recent numerical work by, e.g.,
Holley-Bockelmann et al.\ (2002) suggests that this may not occur on
interesting time scales. Construction of dynamical models of galaxies
with measured black hole masses can provide significant constraints on
the intrinsic shapes and orbital structure, and hence may shed further
light on this problem.

\subsection{Mergers, binary black holes, and gravitational waves}

In the currently popular scenario of hierarchical galaxy formation,
the giant ellipticals with extended cores had their last major merger
many Gyr ago, and since then accreted mostly small lumps. An infalling
galaxy will lose its steep central cusp if the host contains a massive
black hole. This tidally disrupts the cusp, and transforms it into a
nuclear stellar disk of the kind seen by {\it HST}\ in nearby galaxies. The
core itself was formed during the last major merger that presumably
involved two cusped systems, each containing a black hole. Dynamical
friction causes the black holes to form a binary, after which
star-binary interactions remove many of the stars from the central
region, generating an extended low-surface brightness core, or even a
declining central luminosity profile (Lauer). The dynamical
interactions harden the black hole binary, which eventually may reach
the stage where gravitational radiation becomes effective so that
rapid black hole coalescence follows (e.g., Begelman, Blandford \&
Rees 1980). Overall triaxiality of the host galaxy increases the black
hole merging rate because a large fraction of stars are on orbits
that bring them close to the center so that the binary evolution
speeds up (e.g., Yu 2002).

Merritt and Milosavljevi\'c reported on recent progress in this area.
No reliable $N$-body simulations of the complete process of formation of
a core-galaxy via a binary black hole merger of two cusped systems are
available yet, and not much is known about the expected shape of the
resulting core, but this situation should change in the next few
years.  To date, it remains unclear whether black hole coalescence
goes to completion in less than a Hubble time, or whether the binary
stalls. If the process is slow, then binary black holes might be
detectable with high spatial resolution spectroscopy in nearby systems
such as Centaurus A, and they would provide a natural explanation for
the twisted jets seen in some radio galaxies (Merritt \& Ekers 2002).
In this case repeated mergers will cause three-body interactions of
the black holes, which may eject one of them from the galaxy
altogether, and hence generate a population of free-floating massive
black holes hurtling through space. If, on the other hand, black hole
coalescence is fast, then giant ellipticals will contain a
slowly rotating single black hole (Hughes \& Blandford 2003), and many
binary black hole mergers must have occurred in the past. A number of
speakers discussed the exciting possibility of detecting the
gravitational wave signature of these black hole mergers with LISA,
the planned ESA/NASA space observatory to be launched in the next
decade (Armitage, Backer, P.\ Bender, Phinney).

\subsection{Fueling}

The bulk of the fuel for the central engine that powers an active
nucleus must be in gaseous form. While transporting gas inward to
about 100 pc from the nucleus is fairly straightforward, it is not at
all clear how to get it inside 1 pc. Shloshman, Frank \& Begelman's
(1990) bar-in-bar scenario has been updated, but observational
evidence suggests that large-scale bars may not be very important for
fueling.  However, Emsellem presented intriguing kinematic evidence
for inner density waves, perhaps related to central bars, which may
hold the key to efficient gas transport into the nucleus.

Three-dimensional simulations of the structure of nuclear gas continue
to improve. Wada showed a beautiful example in which the gas is
turbulent and filamentary, but is in ordered motion with a significant
velocity dispersion, in qualitative agreement with observations.  The
morphology of his simulated disk resembles that of the celebrated
H$\alpha$ disk in M87 (Ford et al.\ 1994). It will be interesting to
see how well the kinematic properties compare. The turbulent
filamentary structure in the simulated disk is caused by local energy
input which corresponds to an assumed supernova rate of 1 per
year. While this may be appropriate for the peak of activity in
high-luminosity quasars, it is less evident that this would apply to,
say, M87.

\subsection{Accretion physics and feedback}

Blandford reviewed recent progress in our understanding of accretion
physics. The basic paradigm for an active nucleus --- an accretion disk
surrounding a supermassive black hole --- has been agreed for over
30 years, but detailed models have been fiendishly difficult to
construct as they require inclusion of a remarkable range of physical
phenomena in the context of general relativistic
magneto-hydrodynamics.  The goal is to model the morphology and
spectra of individual objects in detail, and to reproduce the full zoo
of the AGN taxonomy, where inclination of the accretion disk to the
line of sight, but also age and accretion mode are parameters that can
be varied (e.g., Umemura).  Theory has identified three viable modes
of accretion. In the ``low'' mode, accretion is adiabatic. This is
likely to apply to ``inactive'' nuclei (low-luminosity AGNs), including
Sgr A* in our own Galaxy. The ``intermediate'' mode may apply to
Seyferts (Heckman) and corresponds to the classical thin disk (Shakura
\& Sunyaev 1973), while the ``high'' mode is again adiabatic, and may be
most relevant for obscured quasars. Amongst the adiabatic models much
work has been done on advection-dominated accretion flows (Narayan \&
Yi 1994), which are radiatively inefficient. There is disagreement as
to whether this ADAF model fits the spectrum of, e.g., Sgr A*. The
theory of the more recent convection-dominated accretion flow (CDAF)
variant is less well developed, and in particular the
magneto-hydrodynamic flow in these models has not been worked out in
full. The adiabatic inflow-outflow solution (ADIOS) of
Blandford \& Begelman (1999) was developed to overcome some of the
difficulties encountered by the ADAF models, but more work on
this is needed as well.

Begelman reminded us that much of the output of an active nucleus is
in kinetic form (outflows, jets), which may inhibit accretion and
affect the surroundings. Some of this feedback may be observable in
radio galaxies, as episodic activity that he argued is
buoyancy-driven, with the energy distributed by effervescent
heating. It will be interesting to work out in more detail how
feedback operates in the formation of the luminous high-$z$
quasars. We have seen that these require rapid and efficient growth of
their central supermassive black hole. It is likely that efficient
cooling is required to bring in enough gas to grow the black hole
efficiently. Statistical evidence from the Sloan Survey suggests that
powerful active nuclei may be accompanied by a significant population
of young and intermediate-age stars, and that the most powerful ones
may well reside in massive young galaxies (Heckman, Kauffmann). The
combined effects of feedback by the growing nucleus, and of the
massive accompanying starburst, may well be reponsible for lifting the
obscuration provided by the dust and gas revealing the quasar for all
the Universe to see.

\section{Black Holes and Galaxy Formation}

It has been clear for some time that nuclear and global properties of
galaxies correlate. This includes the correlation of cusp slope with
total luminosity, the $\mbh-\sigma$ relation, and also the relation
between $\mbh$ and the Sersic index $n$ that characterizes the overall
(rather than the central) luminosity profile (Graham, Erwin).
Understanding these correlations, and establishing which of these is
fundamental, is a key challenge for galaxy formation theories.

The current galaxy formation paradigm is based upon hierarchical
merging, where the dark matter clumps through gravitational
instability in the early Universe, after which the primordial gas
settles in the resulting potential well, in the form of a disk, and
starts forming stars. These produce heavier elements that they return
to the gas at the end of their lives, so that subsequent generations
of stars are increasingly metal rich. The proto-galactic disks
experience frequent interactions, ranging from infall of smaller
clumps that hardly disturb the disk to equal-mass mergers that
result in a spheroidal galaxy, which may reacquire a stellar disk by
further gas infall.

Somerville and Haehnelt suggested that during the first collapse, the
low angular momentum material may already form a bulge, containing a
black hole, and that its properties are related to those of the dark
halo. In this view, the main disk would form/accrete later, and its
properties may have little to do with the black hole, so that
correlations between nuclear and global properties arise naturally for
galaxies with classical bulges. The theorists have various scenarios
to reproduce the $\mbh-\sigma$ relation. The models also
predict the evolution of the black hole mass function, and could in
principle provide the expected LISA signal for merging massive black
holes once the vexing details of black hole binary coalescence have
been sorted out.

The observed range of central cusp slopes in ellipticals, lenticulars,
and classical spiral bulges is larger than predicted by simple
adiabatic growth, but qualitatively consistent with successive merging
of smaller cusped systems with their own black holes. It will be very
interesting to compare detailed numerical simulations of this merging
history with the internal orbital structure derived from dynamical
modeling for individual galaxies.  Burkert reported that the observed
tangential orbital anisotropy is not (yet?) seen in his merger
simulations.

This leaves the disk galaxies with pseudo-bulges, which show little
evidence for nuclear activity. Could they have formed late, as a
stable disk in an isolated dark halo, and without a central black
hole? It has been argued that the pseudo-bulges might be the result of
bar dissolution, but despite earlier reports to the contrary, stellar
bars appear to be robust against dynamical evolution caused by a
central ``point mass,'' as long as its mass is as modest as that of the
observed central star clusters (Carollo, Sellwood, Shen).
Understanding the formation and subsequent internal evolution of these
disks is an area that deserves much more attention.

\section{Challenges}

Rather than summarize this summary, or repeat suggestions for future
work made in the above, I conclude by mentioning two areas where
progress is most needed.

It will be very useful to measure more black hole masses, to (i)
establish the local demography over the full mass range from globular
clusters to giant ellipticals, (ii) settle whether or not the galaxies
with pseudo-bulges contain central black holes, and (iii) calibrate
the secondary and tertiary mass estimators that can then hopefully be
used to estimate reliable black hole masses to high redshift. On the
theoretical side, it will be interesting to see if the measured black
hole masses can help pin down the models for the various active
nuclei, e.g., to establish accretion mode and age.

Pressing projects related to galaxy formation include (i) establishing
what are the fundamental parameters underlying the various
correlations between global and nuclear properties of galaxies, (ii)
understanding the evolution of the black hole mass function, from the
formation of the first seed black holes to the present day, including
the role of binaries, and (iii) working out where the galaxies with
pseudo-bulges fit in.\looseness=-2

\medskip
It is a pleasure to acknowledge stimulating conversations with many of
the participants at this Symposium, to thank Michele Cappellari for
expert assistance during the preparation of the summary talk, and to
thank him, Peter Barthel, Marcella Carollo, Eric Emsellem, Karl
Gebhardt, Luis Ho, and Gijs Verdoes Kleijn for constructive comments
on an earlier version of this manuscript.

\end{document}